 \documentclass{emulateapj}
 \usepackage{apjfonts}

\slugcomment{Not to appear in Nonlearned J., 45.}

\shorttitle{Jet Structures from GRB Polarization}
\shortauthors{Yonetoku et al.}

\begin{document}


\title{Magnetic Structures in Gamma-Ray Burst Jets 
Probed by Gamma-Ray Polarization}


\author{
Daisuke Yonetoku\altaffilmark{1},
Toshio Murakami\altaffilmark{1}, 
Shuichi Gunji\altaffilmark{2}, 
Tatehiro Mihara\altaffilmark{3},
Kenji Toma\altaffilmark{4},
Yoshiyuki Morihara\altaffilmark{1}, 
Takuya Takahashi\altaffilmark{1}, 
Yudai Wakashima\altaffilmark{1}, 
Hajime Yonemochi\altaffilmark{1}, 
Tomonori Sakashita\altaffilmark{1}, 
Noriyuki Toukairin\altaffilmark{2}, 
Hirofumi Fujimoto\altaffilmark{1}, 
and
Yoshiki Kodama\altaffilmark{1}, 
}
\email{yonetoku@astro.s.kanazawa-u.ac.jp}


\altaffiltext{1}{College of Science and Engineering, 
School of Mathematics and Physics,
Kanazawa University, Kakuma, Kanazawa, Ishikawa 920-1192, Japan}
\altaffiltext{2}{Department of Physics, Faculty of Science, 
Yamagata University, 1-4-12, Koshirakawa, Yamagata, Yamagata 990-8560, Japan}
\altaffiltext{3}{Cosmic Radiation Laboratory, RIKEN, 2-1, Hirosawa, 
Wako City, Saitama 351-0198, Japan}
\altaffiltext{4}{Department of Earth and Space Science, 
Osaka University, Toyonaka 560-0043, Japan}

\begin{abstract}
We report polarization measurements in two prompt emissions of
gamma-ray bursts, GRB~110301A and GRB~110721A, observed 
with the Gamma-ray burst polarimeter (GAP) aboard IKAROS 
solar sail mission. We detected linear polarization
signals from each burst with polarization degree of
$\Pi = 70 \pm 22$~\% with statistical significance 
of $3.7~\sigma$ for GRB~110301A, and $\Pi = 84^{+16}_{-28}$~\% 
with $3.3~\sigma$ confidence level for GRB~110721A. 
We did not detect any significant change of polarization angle. 
These two events had shorter durations and dimmer brightness 
compared with GRB~100826A, which showed a significant change 
of polarization angle, as reported in \citet{yonetoku2011b}.
Synchrotron emission model can be consistent with all 
the data of the three GRBs, while photospheric quasi-thermal 
emission model is not favorable. We suggest that magnetic field 
structures in the emission region are 
globally-ordered fields advected from the central engine.
\end{abstract}



\section{Introduction}
Gamma-ray bursts (GRBs) are the most energetic explosions
in the universe. Since the discovery of GRBs in late 1960s, 
spacial distribution (directions and redshifts), 
lightcurves and energy spectra have been observed for 
large amount of GRBs. However, the emission mechanism of 
prompt GRBs, how to release the huge energy in short 
time duration, is still a crucial issue. 
The other characteristic of electro-magnetic wave, 
i.e. polarization in gamma-ray band, is thought to be 
a key to solve the problem. Polarization measurement 
can probe the existence of magnetic fields, 
and/or some kinds of geometrical structures,
which are difficult to be examined from 
time variability and energy spectra.

Several previous works reported marginal detections of 
linear polarization. \citet{coburn2003} reported 
the detection of strong polarization from GRB~021206 
by {\it RHESSI}. However, independent authors analyzed 
the same {\it RHESSI} data, and concluded that 
any polarization signals were not confirmed
\citep{rutledge2004, wigger2004}. 
\citet{kalemci2007, mcglynn2007, gotz2009} reported 
marginal detections of polarization with 
$\sim 2 \sigma$ confidence from GRB~041219 by 
{\it INTEGRAL}-SPI and -IBIS data. 
As the authors themselves discussed that the possibility 
of instrumental systematics could not be completely removed, 
and a part of these results were obviously inconsistent 
with each other \citep[see Section~1 of][]{yonetoku2011b}.
Therefore, further observations with 
reliable instruments are strongly required to confirm 
the existence of polarization in prompt GRBs.

Recently, Gamma-ray burst polarimeter 
\citep[GAP;][]{yonetoku2006, yonetoku2011a, murakami2010}
aboard the {\it IKAROS} solar-power-sail 
\citep{kawaguchi2008, mori2009} measured 
a linear polarization of $\Pi = 27 \pm 11$~\% with
$2.9~\sigma$ confidence level in extremely bright 
GRB~100826A \citep{yonetoku2011b}. It was found that
the polarization angle changes 
during the prompt emission with $3.5~\sigma$ confidence 
level. The GAP detector is designed for the gamma-ray 
polarimetry of prompt GRBs, and also realized 
the quite small systematic uncertainty of 1.8~\% level 
\citep{yonetoku2011a}. This may be the most convincing 
detection of polarization degree so far. But this 
is only one case, we need more sample to establish 
the existence of gamma-ray polarization in prompt GRBs.
In this Letter, we report two more detections of 
gamma-ray polarization in GRB~110301A and GRB~110721A 
observed with {\it IKAROS}-GAP.
We analyzed all data observed by GAP until the end of 2011, 
and found significant polarization signals from these events.

\section{Observations}
\label{sec:observations}
We observed two GRBs, GRB~110301A and GRB~110721A, with 
the Gamma-Ray Burst Polarimeter (GAP) aboard the {\it IKAROS} 
solar power sail. The polarization detection principle of 
GAP is to measure the anisotropic distribution against 
the Compton scattering angle. According to the Klein-Nishina 
cross section, the gamma-ray photons tend to scatter toward 
the vertical direction of polarization vector. 
Therefore the modulation against the azimuth scattering angle 
is observed if the incident gamma-rays are polarized.
The advantages of GAP are the high axial symmetry in shape and 
the high gain uniformity. These are the key to reduce 
the instrumental systematic uncertainties.
Details of the GAP detector are shown in \citet{yonetoku2011a}.

\begin{figure}
\includegraphics[angle=270]{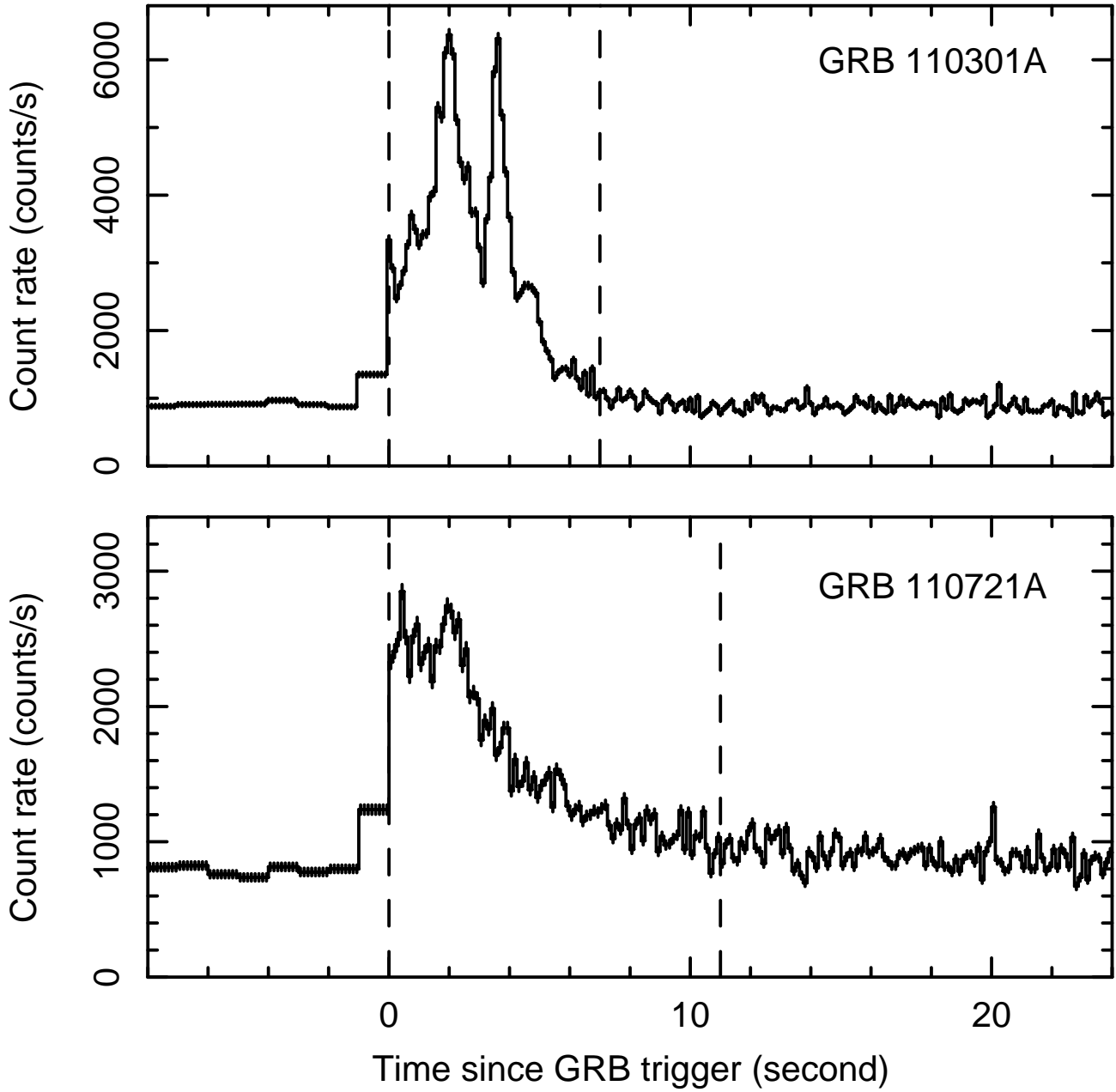}
\caption{Lightcurves of the prompt gamma-ray emission of 
GRB~110301A (top) and GRB~110721A (bottom) detected by GAP. 
The vertical dashed lines indicate the time interval of 
polarization analyses for each burst.
\label{fig1}}
\end{figure}

The GAP detected GRB~110301A on 2011 March 1 at 05:05:34.9 (UT) 
at 0.946 AU away from the Earth. This burst was also detected
by {\it Fermi}-GBM. The coordinate was determined as 
$(\alpha, \delta) = (229.35, +29.40)$ with an uncertainty of 
1.0 degrees in radius \citep{foley11}, which corresponds to 
$48 \pm 1$~deg off-axis from the center of GAP field of view.
Figure~\ref{fig1} (top) shows the lightcurve of GRB~110301A 
observed with GAP in the energy range of 70--300~keV.

According to the time resolved spectral analyses 
by \citet{lu2012}, the spectral peak energy ($E_{p}$) 
slightly changes during the burst, and show hard to soft trend 
from 110~keV to 26~keV. 
Therefore GAP mainly observed the energy range of $E > E_{p}$. 
The energy fluence in 10--1,000~keV is 
$(3.65 \pm 0.03) \times 10^{-5}~{\rm erg~cm^{-2}}$ \citep{foley11}.

GRB~110721A was detected on 2011 July 21 at 04:47:38.9 (UT) 
at 0.699 AU from the Earth. Figure~\ref{fig1} (bottom) 
shows the lightcurve of GRB~110721A. This burst was 
first discovered by {\it Fermi}-GBM and LAT 
\citep{tierney11, vasileiou11}. 
After that, the {\it Swift}-XRT performed the follow-up 
observation of its X-ray afterglow \citep{greiner11, grupe11}. 
The coordinate is precisely measured as 
$(\alpha, \delta) = (333.66, -38.59)$ which corresponds to 
30~deg off-axis. The optical transient 
was also detected by GROND \citep{greiner11} and 
its redshift was measured as $z=0.382$ from two absorption 
lines of Ca{II} with Gemini-South \citep{berger11}. 

The spectral parameters, especially the $E_{p}$ values, 
dramatically change during the burst 
\citep{tierney11, golenetskii2011, lu2012}. 
The $E_{p}$ around the maximum intensity is about 
$E_{p} = 1130^{+550}_{-490}~{\rm keV}$, 
and the one of time integrated spectrum is 
$E_{p} = 393^{+199}_{-104}~{\rm keV}$.
GAP mainly observed in the energy range of $E < E_{p}$.
The energy fluence in 10--1,000~keV is 
$(3.52 \pm 0.03) \times 10^{-5}~{\rm erg~cm^{-2}}$ 
\citep{tierney11}, which is very similar to GRB~110301A.

\section{Data Analyses}
\label{sec:analyses}
\subsection{Average Properties of Polarization}
We analyzed polarization data during the time intervals 
between two dashed lines shown in Figure~\ref{fig1} 
for GRB~110301A and GRB~110721A, respectively. 
GAP obtained the polarization data between -16~s to 176~s 
since the GRB trigger. Since the time durations of 
these GRBs are relatively short, we used the background 
obtained in the same data. 
The net background rate for the polarization data is
$60.0~{\rm counts~s^{-1}}$ for GRB~110301A and
$51.6~{\rm counts~s^{-1}}$ for GRB~110721A.
The total numbers of gamma-ray photons after subtracting 
the background are 1,820 and 1,092 photons for each burst, 
respectively. 

To estimate the systematic uncertainty, we first consider 
the spin rate of {\it IKAROS} spacecraft. The rotation of 
instrument generally reduces the systematic uncertainty 
because the differences of each sensor and 
the geometrical skewness are averaged. 
However, in these case, the time durations of bursts are 
smaller than the period of rotation of {\it IKAROS} 
spacecraft. The spin rate is 1.61~rpm and 0.22~rpm for 
the epoch of GRB~110301A and GRB~110721A, respectively. 
Using the background interval of the data, we created 
the history of background modulation curves with 
the same time interval we analyzed, and confirmed 
each modulation was consistent with constant within 
the statistical error. We confirmed the systematic 
error due to the data analysis of short time duration 
is about $\sigma_{sys,1} = 1.0$~\% of the total polarization 
signals for each bin of both GRBs.

Next, we estimated the systematic uncertainty between 
the detector response calculated by the Geant~4 simulator 
and the experimental data, which is mainly due to 
the off-axis direction of incident gamma-rays. 
We performed several ground experiments 
described in \citet{yonetoku2011b} with 
the proto-flight model of GAP.
We estimated the systematic uncertainty was
$\sigma_{sys,2}= 1.9$~\% of the total polarization 
signals for each bin.

\begin{figure}
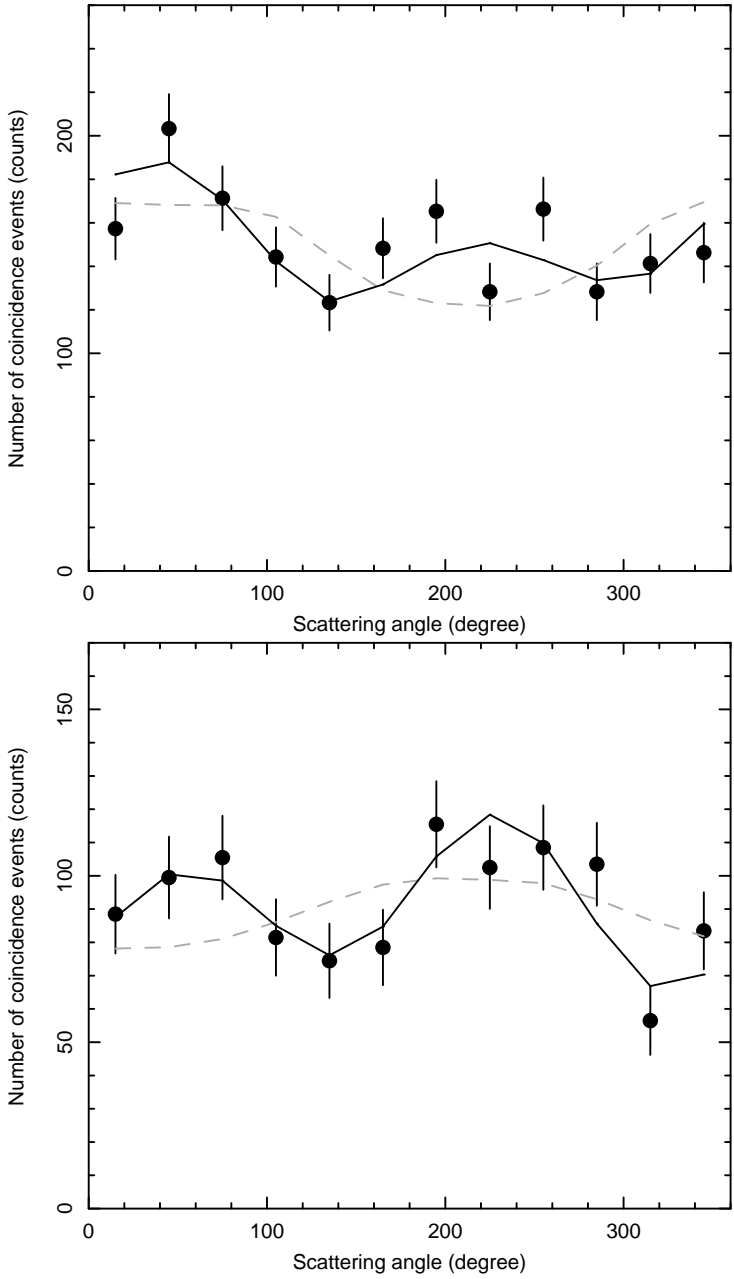

\includegraphics[angle=270,scale=0.5]{fig2a.ps}\\
\includegraphics[angle=270,scale=0.5]{fig2b.ps}
\caption{Number of coincidence gamma-ray photons 
(polarization signals) against the scattering angle of 
GRB~110301A (top) and GRB~110721A (bottom) measured by 
the GAP in 70--300~keV band. 
Black filled circles are the angular distributions of 
Compton scattered gamma-rays after the background 
subtraction. The solid lines are the best fit models 
of each event calculated with the Geant4 Monte-Carlo 
simulations. The model functions of non-polarized cases 
are superposed on each panel for easy comparison. 
\label{fig2}}
\end{figure}

In Figure~\ref{fig2}, we show the modulation curve 
(polarization signals) after subtraction of the background.
The error bars accompanying with the data (filled circles) 
includes not only the statistical error ($\sigma_{stat}$) 
but also the systematic uncertainties described above.
The total errors are calculated as 
$\sigma_{total}^{2} = \sigma_{stat}^{2} + \sigma_{sys,1}^{2} 
+ \sigma_{sys,2}^{2}$ for each bin of polarization data.

The model functions (detector responses) were calculated 
with the Geant~4 simulator considering the spectral 
evolutions reported by \citet{lu2012}, who performed 
spectral analyses for 20 and 14 time intervals of 
GRB~110301A and GRB~110721A, respectively. 
Using their spectral parameters, we simulated  
the model functions for each time interval, and also 
combined into the one with the appropriate weighting 
factor estimated with the brightness histories.

In these analyses, the free parameters are the polarization 
degrees ($\Pi$) and angles ($\phi_p$). 
We simulated the model function with step resolutions of 
5~\% for polarization degree and 5~deg for phase angles. 
In Figure~\ref{fig2}, we show the best fit model with 
solid black lines, and also superposed the non-polarization 
model as the dashed lines on the same panel for easy comparison.
The best fit parameters are $\Pi = 70 \pm 22$~\% and 
$\phi_p = 73 \pm 11$~deg 
with $\chi^{2} = 14.0$ for 10 degree of freedom (d.o.f)
for GRB~110301A, and 
$\Pi = 84^{+16}_{-28}$~\% and $\phi_p = 160 \pm 11$~deg
with $\chi^{2} = 7.3$ for 10 d.o.f 
for GRB~110721A, respectively.
Here the quoted error are at $1~\sigma$ confidence for 
two parameters of interest ($\Pi$ and $\phi_p$), 
and the $\phi_p$ is measured counterclockwise from 
the celestial north.

We show $\Delta \chi^2$ maps in the $(\Pi, \phi_p)$ plane 
in Figure~\ref{fig3}. The white dots are the best fit results,
and we calculate $\Delta \chi^2$ values relative to these points.
The $1 \sigma$, $2 \sigma$ and $3 \sigma$ confidence contours 
for two parameters of interest are shown in the same figures.
The null hypothesis (zero polarization degree) can be ruled 
out with $3.7 \sigma$ (99.98~\%) for GRB~110301A, and 
$3.3 \sigma$ (99.91~\%) confidence levels, respectively.
Although these results have relatively large errors compared 
with the previous GAP result for GRB~100826A 
($\Pi = 27 \pm 11$~\%, $2.9 \sigma$ significance level), 
the polarization degree of these two GRBs may be larger than 
that of GRB~100826A. From these observations, 
we conclude the gamma-ray polarization really exists 
in the prompt GRBs.

\subsection{Time Resolved Polarization Analyses}
\citet{yonetoku2011b} found a significant change of polarization 
angle in very bright GRB~100826A. 
Therefore we also tried to investigate the existence of 
such effect in these two GRBs.

As shown in Figure~\ref{fig1} (top), two main peaks exist 
in the lightcurve of GRB~110301A. We divided the polarization 
data into two parts in time, i.e. $0\;{\rm s} \le t < 3\;{\rm s}$ and 
$3\;{\rm s} \le t < 7\;{\rm s}$.
Here $t$ is the time since trigger. The fitting results are 
$\Pi = 65 \pm 28\;\%$ and $\phi_p = 73 \pm 14\;$deg 
for the first half part, and 
$\Pi = 80^{+20}_{-39}\;\%$ and $\phi_p = 71 \pm 15\;$deg 
for the second half part. 
The significance of polarization detection is
$2.87~\sigma$ and $2.56~\sigma$, respectively.

The brightest part of GRB~110721A lightcurve shows 
rather flatter compared with the standard 
``fast-rise and exponential decay'' of GRBs,
and we divided the polarization data into 
$0\;{\rm s} \le t < 2\;{\rm s}$ and $2\;{\rm s} \le t < 11\;{\rm s}$. 
The best fit results are $\Pi = 81^{+19}_{-40}\;\%$ and 
$\phi_p = 155 \pm 14\;$deg for the first part and
$\Pi = 78^{+22}_{-43}\;\%$ and $\phi_p = 164 \pm 17\;$deg 
for the second part. The significance is $2.55~\sigma$ 
and $2.16~\sigma$, respectively.

We analyzed the polarization data of several time intervals, 
but we could not find any significant change of 
polarization angle and degree. In these two cases, 
we conclude the polarization angles are stable during 
the prompt emission within the error. 
This is quite different properties from GRB~100826A 
by \citet{yonetoku2011b}, and we should consider 
an explainable conditions of emission mechanism
including both polarization properties.

\section{Discussion} \label{discussion}
\label{sec:discussion}
\begin{figure}
\includegraphics[angle=0,scale=0.6]{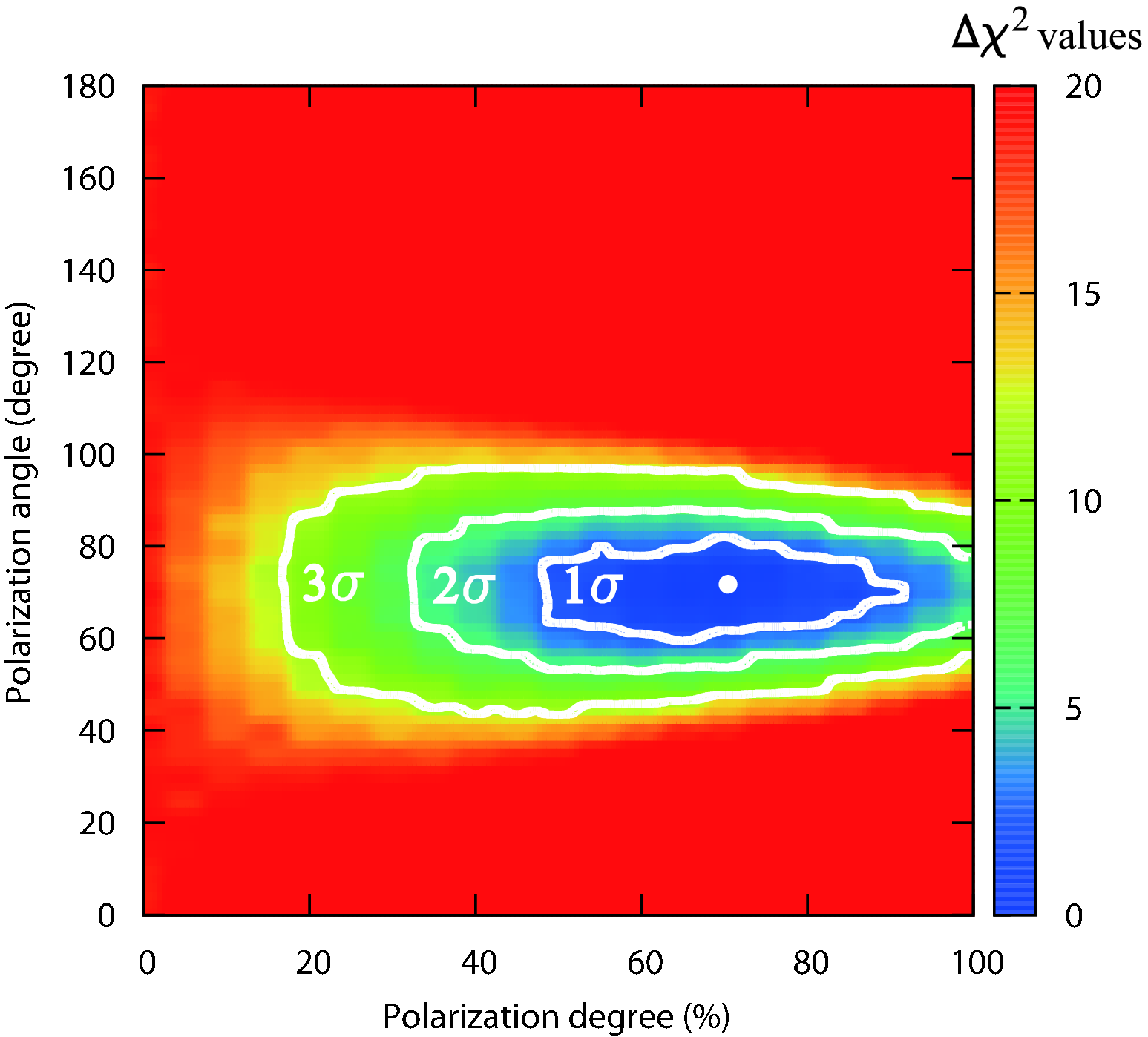}\\
\includegraphics[angle=0,scale=0.6]{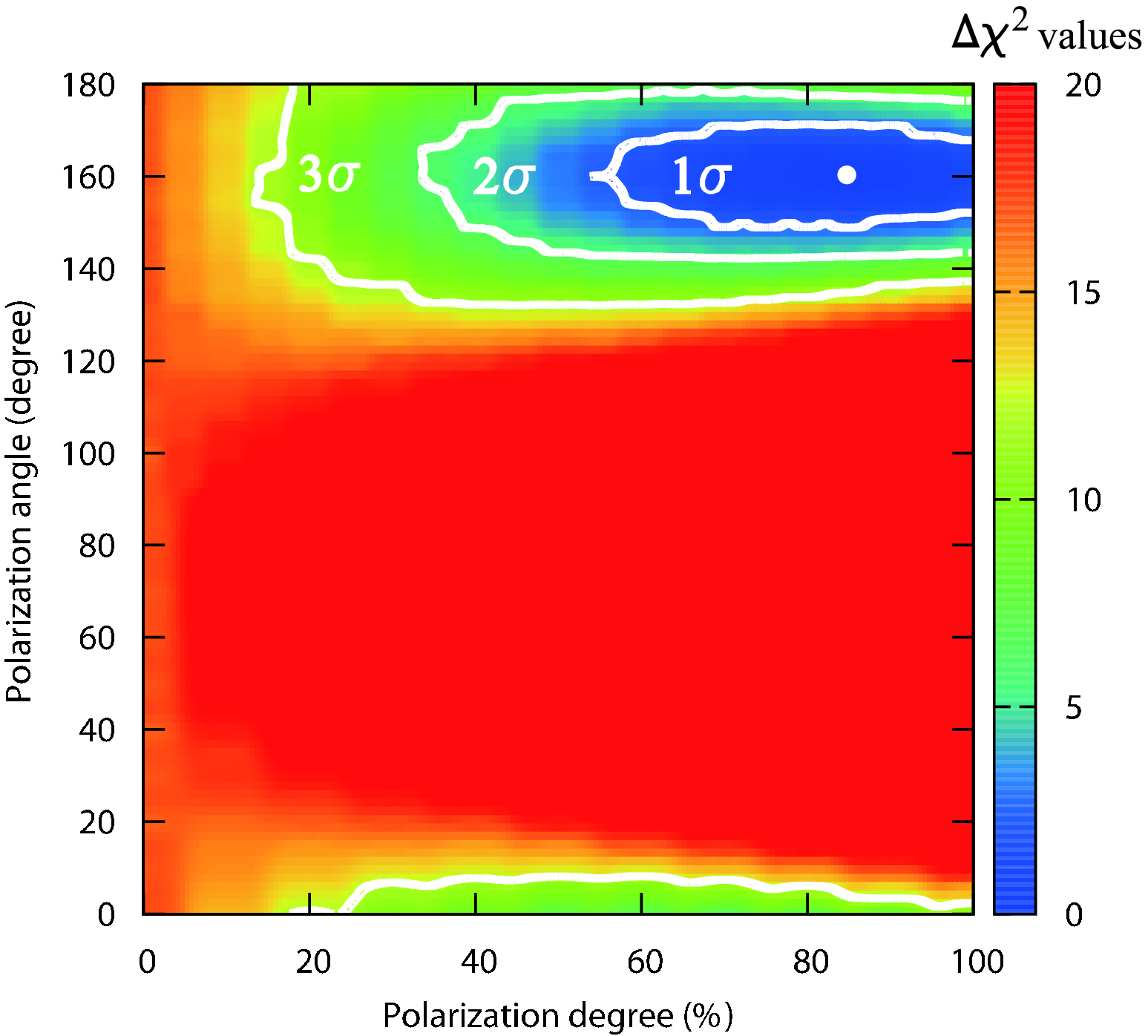}
\caption{Two $\Delta \chi^{2}$ maps of confidence contours 
in the $(\Pi, \phi_p)$ plane for GRB~110301A (top) and
GRB~110721A (bottom). The white dots are the best-fit parameters, 
and we calculate $\Delta \chi^{2}$ values relative to this point. 
A color scale bar along the right side of the contour shows levels 
of $\Delta \chi^{2}$ values. The significances of the existence 
of linear polarization are $3.7~\sigma$ for GRB~110301A and 
$3.3~\sigma$ for GRB110721A, respectively. \label{fig3}}
\end{figure}

Major results of the GAP observations so far are 
simply summarized as follows: 
(1) there are cases with and without a significant change of 
polarization angle (PA), 
and (2) GRB~100826A, with a long duration $T \sim 100\;$s, 
has a PA change, and its polarization degree (PD) is 
$\Pi = 27 \pm 11\%$, while GRB~110301A and GRB~110721A, 
with short durations $T \sim 10\;$s, have no PA change, 
and their PDs are $\Pi \gtrsim 30\%$ at $2 \sigma$ error region
(see Figure~\ref{fig3}). Here we present some theoretical 
implications from these observational results for synchrotron 
and photospheric emission models, following \citet{yonetoku2011b}.

First we discuss synchrotron models with different types of 
magnetic field structure. One type is helical magnetic fields, which 
may be advected from the central engine 
through the jet. Such globally-ordered fields can produce high-PD emission 
\citep{granot03,lyutikov03}.
We call this `SO model'. In this model, the existence of 
a significant PA change implies that emission region consists 
of multiple patches with characteristic angular size $\theta_p$ 
much smaller than the opening angle of the jet $\theta_j$
\citep{yonetoku2011b}.
Relativistic beaming effect only allows us to observe angular 
size of $\sim \Gamma^{-1}$ around the line of sight (LOS), 
where $\Gamma$ is the bulk Lorentz factor of the jet.
In the case of $\Gamma^{-1} \sim \theta_j$, it is natural that 
we see multiple patches with different magnetic field directions, 
and then we observe significant PA changes. 
On the other hand, if $\Gamma^{-1} \ll \theta_j$, we only see 
a limited range of the curved magnetic fields, which leads to 
no significant PA change even for patchy emission. 
The net PD of the latter case can be estimated as $\Pi \sim 40\%$ 
for photon index $\Gamma_B \sim -1$
(whereas synchrotron emission from a point source has 
$\Pi_{\rm max}^{\rm syn} = (-\Gamma_B)/(-\Gamma_B+\frac{2}{3})$)
\citep[see the case of $y_j = 100$ in Figure 2 of][]{toma2009}, 
which is compatible to the observed PD of $\Pi \gtrsim 30\%$ 
from GRB~110301A and GRB~110721A. Therefore, the case of 
$\Gamma^{-1} \sim \theta_j$ could correspond to GRB~100826A, 
while the case of $\Gamma^{-1} \ll \theta_j$ to the other 
two bursts with no PA change.

We may consider an alternative 
model in which GRB jets consist of multiple impulsive shells 
which have globally-ordered transverse magnetic fields with different 
direction for each shell. Let us call this `non-steady SO model'.
It has been recently claimed that such impulsive shells can be 
accelerated to relativistic speeds \citep{granot12}. 
This polarization model may also apply to a scenario in which initial
globally-ordered helical fields get distorted, making different field
directions within the angular size of $\Gamma^{-1}$, as considered
in ICMART model \citep{zhang11}.
In this model, PA changes can naturally occur, 
although it does not necessarily occur when the number of 
shells is small so that $T$ is relatively small.
The net PD will be $\Pi \sim \Pi_{\rm max}^{\rm syn}/\sqrt{N}$,
where $N$ is the number of shells with different field directions. 
This could be as high as $\Pi_{\rm max}^{\rm syn} \sim 60\%$ 
for $\Gamma_B \sim -1$ when $T$ is relatively small, 
which is consistent with the observed PD of $\Pi \gtrsim 30\%$.

Shocks formed in the jet may produce sizable magnetic fields 
with random directions on plasma skin depth scales. 
Synchrotron emission from such fields can have high PD, 
provided that the field directions are not isotropically 
random but random mainly in the plane parallel to the shock 
plane (i.e., perpendicular to the local direction of expansion)
\citep{granot03,nakar03}. 
We call this `SR model'. The linear polarization
directions in the observer frame are symmetric around the LOS. 
Thus, if the emission is patchy, net PD can be nonzero and 
can have PA changes \citep{yonetoku2011b} \citep[see also][]{lazzati09}. 
The characteristic angular size of the patches may be 
hydrodynamically constrained to be $\theta_p \gtrsim \Gamma^{-1}$. 
In this model, the PD of emission from one patch strongly depends 
on the viewing angle of the patch $\theta_v$, 
i.e., the angle between the axis of the patch and the LOS, 
and it is limited as $\Pi \lesssim 40\%$
\citep[see the cases of $y_j \geq 1$ in Figure 3 of][]{toma2009}.
This implies that fine tuning, $\theta_v \sim \theta_p + \Gamma^{-1}$, 
is required to have $\Pi > 30\%$.
Furthermore, the bursts we observed are all very bright, which suggests
that some patches are seen with $\theta_v \lesssim \theta_p$, for which
$\Pi \ll 40\%$. This suggests that the SR model is
not favorable to explain the observed PD of $\Pi \gtrsim 30\%$.

Internal shocks may also produce strong magnetic fields with random
directions on hydrodynamic scales \citep{inoue11,gruzinov99}. We 
call this `SH model', which produces net PD as 
$\Pi \sim \Pi^{\rm syn}_{\rm max}/\sqrt{N}$, where $N$ is the number
of independent patches with coherent magnetic field in the 
observable angular size $\sim \Gamma^{-1}$,
and can naturally lead to PA changes.
Unlike the SR models, the emission from patches seen with small 
$\theta_v$ has high PD, so that this model is in agreement with
the high brightness of the bursts. 
However, recent MHD simulations of internal shocks with
initial density fluctuations imply $N \sim 10^3$ \citep{inoue11}, which
cannot explain the observed PD of $\Pi \gtrsim 30\%$.

Lastly, we discuss photospheric emission (Ph) model. This model
assumes that the emission at $E \gtrsim E_p$ 
is the quasi-thermal radiation from the photosphere. 
The emission at $E < E_p$ might be a
superposition of many of the quasi-thermal components with
different temperatures \citep{ryde10,toma11} or contribution
from synchrotron emission \citep{vurm11}. The quasi-thermal
radiation can have high PD when the radiation energy is smaller 
than the baryon kinetic energy at the photosphere \citep{beloborodov11}. 
The linear polarization directions in the observer frame are 
symmetric around the LOS, same as the SR model. 
The PD of emission from a given point may be 
determined by the brightest emission, coming from just below 
the photosphere, which can have 
$\Pi \leq \Pi_{\rm max}^{\rm qt} \sim 40\%$ \citep{beloborodov11}.
The PD of emission from one patch, whose angular size is 
hydrodynamically constrained to be $\theta_p \gtrsim \Gamma^{-1}$, 
will reduce to $\Pi \lesssim 30\%$ (in the same way as in the SR model).
The patches with small viewing angle $\theta_v$ have $\Pi \ll 30\%$.
Therefore, the Ph model requires very fine tuning of $\theta_v$ to 
reproduce the PD of GRB~110301A, mainly at
$E > E_p$ (i.e., dominated by the quasi-thermal component).

To summarize, (1) the SR, SH and Ph models are not favorable to 
reproduce all of our polarimetric observation results of the three GRBs,
and (2) the SO and non-steady SO could explain all of them.

Recently, early optical polarization from the forward shock
of a GRB afterglow has been detected as $\Pi \simeq 10.4 \pm 2.5$~\%
(at $t = 149-706\;$s) for GRB~091208B \citep{uehara12}. 
If the magnetic field directions in the emission region are 
random on plasma skin depth scales 
(i.e., the SR model for a shock produced in the circumburst medium), 
the observed polarization reaches a maximum value around 
the jet break time ($\sim 1\;$day) \citep{lazzati2006}.
The measurement of the early polarization, higher than the typical
observed late-time polarization of $\sim 1-3$~\% 
\citep[at $t \sim 1\;$day;][]{covino2004},
thus disfavors this model. This result is consistent with 
our argument against the SR model for a shock produced within the jet.

The {\it Fermi} satellite team has suggested that GRB 110721A has 
a blackbody component with temperature ranging in the $\sim 10-100\;$keV
\citep{axelsson12}, which almost coincide with the GAP range $70-300\;$keV.
However, their fitting model of the integrated spectrum (their Figure~2) 
includes the blackbody flux as only $\sim 1/5{\rm th}$ of the total flux 
in the GAP range. Thus reduction of PD by addition of the blackbody
component is small. The polarization degree and angle in the GAP range
is practically determined by those of the non-thermal component, which
could be synchrotron emission.

The SO models assume globally-ordered fields in the emission region.
On the other hand, observations have suggested that 
prompt emission has very high efficiency 
\citep[even $>90\%$ for some bursts;][]{zhang07,ioka06},
which means that energy dissipation, usually involving field distortion, occur globally. 
Reconciling high PD with high efficiency looks a dilemma, which will have to be
resolved in more quantitative modeling.

\acknowledgments
We thank the referee for several useful comments.
This work is supported by the Grant-in-Aid for 
Young Scientists (S) No.20674002 (DY), 
Young Scientists (A) No.18684007 (DY),
JSPS Research Fellowships for Young Scientists No.231446 (KT), and 
also supported by the Steering Committee for Space Science 
at ISAS/JAXA of Japan.

\clearpage

\clearpage

\clearpage

\end{document}